\documentclass[a4paper,12pt]{article}
\usepackage{amssymb}
\usepackage{amsmath}
\usepackage{color}
\usepackage[unicode,bookmarks,bookmarksopen,bookmarksopenlevel=2,colorlinks,linkcolor=blue,citecolor=green]{hyperref}

\input{tcilatex}
\begin{document}

\title{\textbf{Lagrangian and Hamiltonian structures for the Constant
Astigmatism Equation}}
\author{Maxim V.~Pavlov$^{1}$, Sergey A. Zykov$^{2}$ \\
$^{1}$Sector of Mathematical Physics,\\
Lebedev Physical Institute of Russian Academy of Sciences,\\
Moscow, Leninskij Prospekt, 53, Russia\\
$^{2}$Mathematical Institute in Opava, Silesian University in Opava,\\
Na Rybn\'{\i}\v{c}ku 1, 746 01 Opava, Czech Republic}
\date{}
\maketitle

\begin{abstract}
In this paper we found a Lagrangian representation and corresponding
Hamiltonian structure for the constant astigmatism equation. Utilizing this
Hamiltonian structure and extra conservation law densities we construct a
first evolution commuting flow of the third order. Also, we apply the
recursion operator and present a second Hamiltonian structure. This
bi-Hamiltonian structure allows to replicate infinitely many local commuting
flows and corresponding local conservation law densities.
\end{abstract}

\tableofcontents

\section{Introduction}

Plenty of integrable equations were found in classical differential
geometry. One of them is the famous Bonnet equation also known as the
sin-Gordon equation. This equation expresses angles between asymptotic
directions of surfaces of negative constant Gaussian curvature. Recently,
interest to the Bonnet equation was renew due to a successful search of
integrable cases of Weingarten surfaces. The equation describing surfaces of
constant astigmatism%
\begin{equation}
u_{tt}+\left( \frac{1}{u}\right) _{xx}+2=0  \label{asti}
\end{equation}%
was considered in a set of papers (see detail in \cite{MM}, \cite{MM1}, \cite%
{MM2}, \cite{MM3}). Also, a transformation between the Bonnet equation and (%
\ref{asti}) was found. However this transformation is very sophisticated
(see, for instance, \cite{MM3}). It is not so easy to recompute solutions
and Hamiltonian structures from the Bonnet equation to (\ref{asti}). By this
reason, we construct independently the Lagrangian, corresponding Hamiltonian
structure, first evolution commuting flow of the third order and a second
Hamiltonian structure of (\ref{asti}) in this paper. The inverse
transformation (from (\ref{asti}) back to the Bonnet equation) is not so
complicated. We believe that our results can be effectively utilized in the
theory of the Bonnet equation.

\section{Lagrangian and Hamiltonian Structure}

The Lagrangian%
\begin{equation}
S=\int \left( \frac{1}{2}\Omega _{xx}\Omega _{tt}-f(\Omega ,\Omega
_{x},\Omega _{xx},...)\right) dxdt  \label{Lagrange}
\end{equation}%
determines the Euler--Lagrange equation%
\begin{equation}
\Omega _{xxtt}=\frac{\delta \mathbf{F}}{\delta \Omega },  \label{Euler}
\end{equation}%
where $\mathbf{F}=\int f(\Omega ,\Omega _{x},\Omega _{xx},...)dx$.
Obviously, two local conservation laws (of the energy and of the momentum)
can be obtained (due to E. Noether's Theorem) from the energy-momentum
tensor. For instance, the conservation law of the momentum is%
\begin{equation*}
(\Omega _{xx}\Omega _{xt})_{t}=\left( \frac{1}{2}\Omega _{xt}^{2}+\Omega
_{x}\Omega _{xtt}-G\right) _{x},
\end{equation*}%
where $G_{x}=\frac{\delta \mathbf{F}}{\delta \Omega }\Omega _{x}$, while the
conservation law of the energy is%
\begin{equation*}
\left( \frac{1}{2}\Omega _{xt}^{2}+f(\Omega ,\Omega _{x},\Omega
_{xx},...)\right) _{t}=(\Omega _{t}\Omega _{xtt}+Q)_{x},
\end{equation*}%
where $Q_{x}=-\frac{\delta F}{\delta \Omega }\Omega _{t}+\frac{\partial f}{%
\partial \Omega }\Omega _{t}+\frac{\partial f}{\partial \Omega _{x}}\Omega
_{xt}+\frac{\partial f}{\partial \Omega _{xx}}\Omega _{xxt}+...$

If, for instance, $f(\Omega ,\Omega _{x},\Omega _{xx})$, then%
\begin{equation*}
(\Omega _{xx}\Omega _{xt})_{t}=\left( \frac{1}{2}\Omega _{xt}^{2}+\Omega
_{x}\Omega _{xtt}-f+\frac{\partial f}{\partial \Omega _{x}}\Omega
_{x}-\left( \frac{\partial f}{\partial \Omega _{xx}}\right) _{x}\Omega _{x}+%
\frac{\partial f}{\partial \Omega _{xx}}\Omega _{xx}\right) _{x},
\end{equation*}%
\begin{equation*}
\left( \frac{1}{2}\Omega _{xt}^{2}+f(\Omega ,\Omega _{x},\Omega
_{xx})\right) _{t}=\left( \Omega _{xtt}\Omega _{t}+\frac{\partial f}{%
\partial \Omega _{x}}\Omega _{t}-\left( \frac{\partial f}{\partial \Omega
_{xx}}\right) _{x}\Omega _{t}+\frac{\partial f}{\partial \Omega _{xx}}\Omega
_{xt}\right) _{x},
\end{equation*}%
while Euler--Lagrange equation (\ref{Euler}) can be written as a Hamiltonian
system%
\begin{equation*}
\Omega _{t}=\partial _{x}^{-1}\frac{\delta \mathbf{H}}{\delta w},\text{ \ \ }%
w_{t}=\partial _{x}^{-1}\frac{\delta \mathbf{H}}{\delta \Omega },\text{ \ }%
\mathbf{H}=\int (f(\Omega ,\Omega _{x},\Omega _{xx})+\frac{1}{2}w^{2})dx,
\end{equation*}%
where $w=\Omega _{xt}$.

In this paper we consider just the case $f(\Omega ,\Omega _{x},\Omega
_{xx})=-2\Omega -\ln \Omega _{xx}$. Corresponding Euler--Lagrange equation (%
\ref{Euler}) is nothing but constant astigmatism equation (\ref{asti}),
where $u=\Omega _{xx}$. Thus, constant astigmatism equation (\ref{asti})
possesses the local Lagrangian representation%
\begin{equation}
S=\int \left( \frac{1}{2}\Omega _{xx}\Omega _{tt}+\ln \Omega _{xx}+2\Omega
\right) dxdt,  \label{act}
\end{equation}%
two local conservation laws (the momentum and the energy, respectively):%
\begin{equation*}
(\Omega _{xx}\Omega _{xt})_{t}=\left( \frac{1}{2}\Omega _{xt}^{2}+\Omega
_{x}\Omega _{xtt}+2\Omega -\frac{\Omega _{x}\Omega _{xxx}}{\Omega _{xx}^{2}}%
+\ln \Omega _{xx}\right) _{x},
\end{equation*}%
\begin{equation*}
\left( \frac{1}{2}\Omega _{xt}^{2}-2\Omega -\ln \Omega _{xx}\right)
_{t}=\left( \Omega _{xtt}\Omega _{t}-\frac{\Omega _{xxx}}{\Omega _{xx}^{2}}%
\Omega _{t}-\frac{\Omega _{xt}}{\Omega _{xx}}\right) _{x}
\end{equation*}%
and non-local Hamiltonian structure

\begin{equation}
\Omega _{t}=\partial _{x}^{-1}\frac{\delta \mathbf{H}}{\delta w},\text{ \ \ }%
w_{t}=\partial _{x}^{-1}\frac{\delta \mathbf{H}}{\delta \Omega },
\label{ham}
\end{equation}%
where the Hamiltonian $\mathbf{H}=\int (\frac{1}{2}w^{2}-2\Omega -\ln \Omega
_{xx})dx$ and the momentum $\mathbf{P}=\int \Omega _{xx}wdx$.

\textbf{Remark}: Under the substitution $u=\Omega _{xx}$, the above
non-local Hamiltonian structure assumes a local form%
\begin{equation}
u_{y}=\partial _{x}\frac{\delta \mathbf{H}}{\delta w},\text{ \ \ }%
w_{y}=\partial _{x}\frac{\delta \mathbf{H}}{\delta u},  \label{mod}
\end{equation}%
where the momentum $\mathbf{P}=\int uwdx$ still is local, but the
Hamiltonian $\mathbf{H}=\int (\frac{1}{2}w^{2}-\ln u-2\Omega )dx$ is
essentially nonlocal.

Also constant astigmatism equation (\ref{asti}) has extra two conservation
laws (see \cite{MM3})%
\begin{equation*}
\partial _{t}\sqrt{4u+\left( \frac{u_{x}}{u}\pm u_{t}\right) ^{2}}=\pm
\partial _{x}\sqrt{\frac{4}{u}+\left( \frac{u_{x}}{u^{2}}\pm \frac{u_{t}}{u}%
\right) ^{2}}.
\end{equation*}%
Thus, one can construct a third order symmetry (cf. (\ref{mod}))%
\begin{equation}
u_{y}=\partial _{x}\frac{\delta \mathbf{\tilde{H}}}{\delta w},\text{ \ \ }%
w_{y}=\partial _{x}\frac{\delta \mathbf{\tilde{H}}}{\delta u},\text{ \ }%
\mathbf{\tilde{H}}=\int \sqrt{4u+\left( \frac{u_{x}}{u}\pm w_{x}\right) ^{2}}%
dx.  \label{new}
\end{equation}

\textbf{Remark}: Any higher commuting flow to constant astigmatism equation (%
\ref{asti}) also can be written via the same function $\Omega $ only.
Indeed, taking into account that $u=\Omega _{xx}$ and $w=\Omega _{xt}$,
evolution system (\ref{new}) reduces to two three dimensional equations%
\begin{equation}
\Omega _{xxt}+2\Omega _{y}\sqrt{\frac{\Omega _{xx}}{1-\Omega _{y}^{2}}}\pm 
\frac{\Omega _{xxx}}{\Omega _{xx}}=0,\text{ \ \ }\Omega _{yt}=\sqrt{\frac{%
1-\Omega _{y}^{2}}{\Omega _{xx}}}\pm \frac{\Omega _{xy}}{\Omega _{xx}}.
\label{tri}
\end{equation}%
The compatibility condition $(\Omega _{xxt})_{y}=(\Omega _{yt})_{xx}$ leads
to the single equation%
\begin{equation*}
\left( \frac{\Omega _{xxx}}{\Omega _{xx}}\pm 2\Omega _{y}\sqrt{\frac{\Omega
_{xx}}{1-\Omega _{y}^{2}}}\right) _{y}+\left( \frac{\Omega _{xy}}{\Omega
_{xx}}\pm \sqrt{\frac{1-\Omega _{y}^{2}}{\Omega _{xx}}}\right) _{xx}=0,
\end{equation*}%
which is nothing but an Euler--Lagrange equation associated with the local
Lagrangian representation (cf. (\ref{act}))%
\begin{equation*}
\tilde{S}=\int [\Omega _{xy}\ln \Omega _{xx}\pm 2\sqrt{\Omega _{xx}(1-\Omega
_{y}^{2})}]dxdy.
\end{equation*}

Meanwhile, one can express $\Omega _{y}$ from the first equation in (\ref%
{tri}) and substitute it into the second equation in (\ref{tri}). This gives
again the constant astigmatism equation.

\section{Bi-Hamiltonian structure}

In this Section we present a second Hamiltonian structure for constant
astigmatism equation (\ref{asti}) and its hierarchy.

Infinitely many symmetries (here $t^{k}$ are group parameters)%
\begin{equation}
u_{t^{k+1}}=(-u_{t}\partial _{x}^{-1}+u_{x}\partial _{x}^{-2}\partial
_{t}+2u\partial _{x}^{-1}\partial _{t})u_{t^{k}},\text{ }k=1,2,...
\label{sym}
\end{equation}%
of constant astigmatism equation (\ref{asti}) are connected to each other by
the recursion operator (found by A. Sergyeyev, see in \cite{MM3})

\begin{equation*}
R=-u_{t}\partial _{x}^{-1}+u_{x}\partial _{x}^{-2}\partial _{t}+2u\partial
_{x}^{-1}\partial _{t}.
\end{equation*}

However, for construction of a second Hamiltonian structure, this recursion
operator at first should be rewritten in a matrix form, because constant
astigmatism equation (\ref{asti}) is a two component system. Indeed,
introducing the field variable $q$ such that $u_{t}=q$, (\ref{asti}) takes
an evolution form%
\begin{equation}
u_{t}=q,\text{ \ }q_{t}=-\left( \frac{1}{u}\right) _{xx}-2.  \label{b}
\end{equation}%
Thus, differentiating (\ref{sym}) with respect to $t$, one can obtain%
\begin{equation*}
q_{t^{k+1}}=-q_{t}\partial _{x}^{-1}u_{t^{k}}+q_{x}\partial
_{x}^{-2}q_{t^{k}}+q\partial _{x}^{-1}q_{t^{k}}+u_{x}\partial
_{x}^{-2}(q_{t})_{t^{k}}+2u\partial _{x}^{-1}(q_{t})_{t^{k}}.
\end{equation*}%
Finally, taking into account (\ref{b}), we eliminate derivatives with
respect to $t$. This yields a desirable relationship%
\begin{equation*}
u_{t^{k+1}}=-q\partial _{x}^{-1}u_{t^{k}}+(2u\partial
_{x}^{-1}+u_{x}\partial _{x}^{-2})q_{t^{k}},
\end{equation*}%
\begin{equation*}
q_{t^{k+1}}=\left( \frac{2}{u}\partial _{x}-3\frac{u_{x}}{u^{2}}+\left[
\left( \frac{1}{u}\right) _{xx}+2\right] \partial _{x}^{-1}\right)
u_{t^{k}}+(q\partial _{x}^{-1}+q_{x}\partial _{x}^{-2})q_{t^{k}}.
\end{equation*}%
Thus, under the potential substitution $q=w_{x}$, the above transformation
of symmetries can be written in the matrix form%
\begin{equation*}
\left( 
\begin{array}{c}
u \\ 
w%
\end{array}%
\right) _{t^{k+1}}=\hat{R}\left( 
\begin{array}{c}
u \\ 
w%
\end{array}%
\right) _{t^{k}}=\left( 
\begin{array}{cc}
-w_{x}\partial _{x}^{-1} & 2u+u_{x}\partial _{x}^{-1} \\ 
2\frac{1}{u}-\frac{u_{x}}{u^{2}}\partial _{x}^{-1}+2\partial _{x}^{-2} & 
w_{x}\partial _{x}^{-1}%
\end{array}%
\right) \left( 
\begin{array}{c}
u \\ 
w%
\end{array}%
\right) _{t^{k}}.
\end{equation*}

Since, any local symmetry to (\ref{asti}) has the same local Hamiltonian
structure%
\begin{equation*}
u_{t^{k}}=\partial _{x}\frac{\delta \mathbf{H}_{k}}{\delta w},\text{ \ \ }%
w_{t^{k}}=\partial _{x}\frac{\delta \mathbf{H}_{k}}{\delta u},
\end{equation*}%
we obtain automatically a second Hamiltonian structure%
\begin{equation*}
u_{t^{k}}=\partial _{x}\frac{\delta \mathbf{H}_{k}}{\delta w}=(2u\partial
_{x}+u_{x})\frac{\delta \mathbf{H}_{k-1}}{\delta u}-w_{x}\frac{\delta 
\mathbf{H}_{k-1}}{\delta w},
\end{equation*}%
\begin{equation*}
w_{t^{k}}=\partial _{x}\frac{\delta \mathbf{H}_{k}}{\delta u}=w_{x}\frac{%
\delta \mathbf{H}_{k-1}}{\delta u}+\left( \frac{2}{u}\partial _{x}-\frac{%
u_{x}}{u^{2}}+2\partial _{x}^{-1}\right) \frac{\delta \mathbf{H}_{k-1}}{%
\delta w}.
\end{equation*}

If, for instance, we start from third order evolution system (\ref{asti})%
\begin{equation*}
u_{t^{1}}=\partial _{x}\frac{\delta \mathbf{H}_{1}}{\delta w},\text{ \ \ }%
w_{t^{1}}=\partial _{x}\frac{\delta \mathbf{H}_{1}}{\delta u},\text{ \ }%
\mathbf{H}_{1}=\int \sqrt{4u+\left( \frac{u_{x}}{u}\pm w_{x}\right) ^{2}}dx,
\end{equation*}%
then the next commuting flow%
\begin{equation*}
u_{t^{2}}=\partial _{x}\frac{\delta \mathbf{H}_{2}}{\delta w}=(2u\partial
_{x}+u_{x})\frac{\delta \mathbf{H}_{1}}{\delta u}-w_{x}\frac{\delta \mathbf{H%
}_{1}}{\delta w},
\end{equation*}%
\begin{equation*}
w_{t^{2}}=\partial _{x}\frac{\delta \mathbf{H}_{2}}{\delta u}=w_{x}\frac{%
\delta \mathbf{H}_{1}}{\delta u}+\left( \frac{2}{u}\partial _{x}-\frac{u_{x}%
}{u^{2}}+2\partial _{x}^{-1}\right) \frac{\delta \mathbf{H}_{1}}{\delta w}
\end{equation*}%
also will be again a local symmetry.

Thus, infinitely many local commuting flows constructed from this
bi-Hamiltonian structure can be utilized for description of multi-phase
solutions for constant astigmatism equation (\ref{asti}).

\textbf{Remark}: The second Hamiltonian structure%
\begin{equation*}
\left( 
\begin{array}{c}
u \\ 
w%
\end{array}%
\right) _{t^{k}}=\left( 
\begin{array}{cc}
2u\partial _{x}+u_{x} & -w_{x} \\ 
w_{x} & \frac{2}{u}\partial _{x}-\frac{u_{x}}{u^{2}}+2\partial _{x}^{-1}%
\end{array}%
\right) \left( 
\begin{array}{c}
\frac{\delta \mathbf{H}_{k-1}}{\delta u} \\ 
\frac{\delta \mathbf{H}_{k-1}}{\delta w}%
\end{array}%
\right) .
\end{equation*}%
is nonlocal. This is a linear combination of local Hamiltonian structure of
the Dubrovin--Novikov type and a pure nonlocal part. Such Hamiltonian
structures were investigated in \cite{Fer}. In general $N$ component case,
corresponding Hamiltonian operators have the form ($\alpha \neq 0$ is an
arbitrary constant)%
\begin{equation*}
A^{ij}=g^{ij}\partial _{x}-g^{is}\Gamma _{sk}^{j}u_{x}^{k}+\alpha
f^{i}\partial _{x}^{-1}f^{j},
\end{equation*}%
where $g^{ij}(\mathbf{u})$ is a nondegenerate symmetric metric and $\Gamma
_{sk}^{j}$ are Christoffel symbols of Levi-Civita connection, while $f^{i}$
are components of isometry $f^{i}\partial /\partial u^{i}$, which satisfies
some special conditions. In this Section, we presented a first example of
integrable systems which is equipped by pair of Hamiltonian operators%
\begin{equation}
A^{ij}=g^{ij}\partial _{x}-g^{is}\Gamma _{sk}^{j}u_{x}^{k},\text{ \ }\tilde{A%
}^{ij}=\tilde{g}^{ij}\partial _{x}-\tilde{g}^{is}\tilde{\Gamma}%
_{sk}^{j}u_{x}^{k}+\alpha f^{i}\partial _{x}^{-1}f^{j}.  \label{c}
\end{equation}%
For the integrable hierarchy of constant astigmatism equation we choose $%
u^{1}=u,u^{2}=w$. Then $\alpha =2$ and%
\begin{equation*}
g^{ik}=\left( 
\begin{array}{cc}
0 & 1 \\ 
1 & 0%
\end{array}%
\right) ,\text{ \ }\tilde{g}^{ik}=2\left( 
\begin{array}{cc}
u & 0 \\ 
0 & \frac{1}{u}%
\end{array}%
\right) ,\text{ \ }f^{i}=\left( 
\begin{array}{c}
0 \\ 
1%
\end{array}%
\right) ,
\end{equation*}%
while $\Gamma _{sk}^{j}=0$ and $\tilde{\Gamma}_{11}^{1}=-\tilde{\Gamma}%
_{12}^{2}=\frac{1}{2u},\tilde{\Gamma}_{22}^{1}=\frac{1}{2}u,\tilde{\Gamma}%
_{21}^{2}=-\frac{1}{2}u^{-1},\tilde{\Gamma}_{12}^{1}=\tilde{\Gamma}_{11}^{2}=%
\tilde{\Gamma}_{21}^{1}=\tilde{\Gamma}_{22}^{2}=0$.

The classification of integrable systems determined by pairs of compatible
Poisson brackets (\ref{c}) will be published elsewhere.

\section*{Acknowledgement}

We thank Eugene Ferapontov and Sergey Tsarev for their stimulating and
clarifying discussions.

SAZ was supported by GA\v{C}R under the project P201/11/0356. MVP had a
financial support under the project CZ.1.07/2.3.00/20.0002. MVP is grateful
to professor Michal Marvan for verification of the recursion operator and
for the warm hospitality in the Mathematical Institute in Opava where some
part of this work has been done.

\end{document}